\PassOptionsToPackage{usenames,dvipsnames}{xcolor}
\PassOptionsToPackage{tracking=true,kerning=true,spacing=true,factor=1100,stretch=10,shrink=10}{microtype}

\documentclass[acmtog,screen=true]{acmart} 

\acmJournal{TOG}

\setcopyright{acmlicensed}

\usepackage[l2tabu,orthodox]{nag}

\usepackage[T1]{fontenc}
\usepackage[utf8]{inputenc}

\usepackage{amsthm}
\usepackage{amssymb}
\usepackage{mathtools} 
\usepackage{textcomp}
\usepackage[mathscr]{eucal}
\usepackage{stmaryrd}
\usepackage{physics} 
\usepackage{microtype}

\usepackage{fancyhdr}
\usepackage{ifthen}
\usepackage{cleveref}
\usepackage{csquotes}
\usepackage{calc}
\usepackage[perpage]{footmisc}

\usepackage{alltt}
\usepackage[ruled,vlined,onelanguage,linesnumbered]{algorithm2e}

\usepackage{array}
\usepackage{booktabs}
\usepackage{multirow}

\usepackage{float}
\usepackage{wrapfig} 
\usepackage{graphicx}
\usepackage{subcaption} 
\usepackage{overpic}
\usepackage{tikz}
\usetikzlibrary{decorations.pathmorphing,patterns,shadows,calc}

\usepackage{fixltx2e}

\newcommand{\todo}[1]{\textcolor{red}{TODO: #1 $\qed$}}
\newcommand{\revised}[1]{#1}
\newcommand{\minor}[1]{\textcolor{blue}{#1}}

\definecolor{ReviewColor}{rgb}{0.6,0.15,0.05}

\renewcommand{\revised}[1]{#1}
\renewcommand{\minor}[1]{#1}

\renewcommand{\todo}[1]{}

\hypersetup{
	unicode=true,
	colorlinks=true,
	citecolor=black, 
	linkcolor=NavyBlue, 
	urlcolor=NavyBlue, 
}

\definecolor{forestgreen}{rgb}{0.13,0.54,0.13}

\SetCommentSty{algocomment}

\newcolumntype{M}[1]{>{\centering\arraybackslash}m{#1}}

\definecolor{Stretch_1}{rgb}{0,0.85,0.08}
\definecolor{Stretch_3}{rgb}{0.85,0.25,0}
\definecolor{Stretch_5}{rgb}{0.16,0,0.85}

\let\originalleft\left
\let\originalright\right
\renewcommand{\left}{\mathopen{}\mathclose\bgroup\originalleft}
\renewcommand{\right}{\aftergroup\egroup\originalright}

\renewcommand{\geq}{\geqslant}
\renewcommand{\leq}{\leqslant}

\begin{document}

\title{Colored Fused Filament Fabrication}

\newcommand{\mfx}{Universit\'{e} de Lorraine, CNRS, Inria, LORIA}

\author{Haichuan Song}
\author{Jonàs Mart\'{i}nez}
\author{Pierre Bedell}
\author{Noémie Vennin}
\author{Sylvain Lefebvre}
\affiliation{
	\postcode{F-54000}
	\city{Nancy}
	\state{France}
	\institution{\mfx}
}

\begin{abstract}
    Fused filament fabrication is the method of choice for printing 3D models
    at low cost and is the de-facto standard for hobbyists, makers, and schools.
    Unfortunately, filament printers cannot truly reproduce colored objects.
    The best current techniques rely on a form of dithering exploiting 
    occlusion, that was only demonstrated for shades of two base colors and 
    that behaves differently depending on surface slope.
    
    We explore a novel approach for 3D printing colored objects,
    capable of creating controlled gradients of varying
    sharpness. Our technique exploits off-the-shelves nozzles that are 
    designed to mix multiple filaments in a small melting chamber, 
    obtaining intermediate colors once the mix is stabilized.
    
    We apply this property to produce color gradients. 
    We divide each input layer into a set of strata, each having
    a different constant color. By locally changing the thickness of
    the stratum, we change the perceived color at a given 
    location. By optimizing the choice of colors of each stratum, we 
    further improve quality and allow the use of different numbers of 
    input filaments.
    
    We demonstrate our results by building a functional color printer using
    low cost, off-the-shelves components. Using our tool a user can paint a 3D
    model and directly produce its physical counterpart, using any material
    and color available for fused filament fabrication.
    
\end{abstract}

%
%


\thanks{
	This work is supported by ERC grant ShapeForge (StG-2012-307877).
}

\begin{teaserfigure}
    \centering
    \raisebox{0.0cm}{\includegraphics[height=5cm]{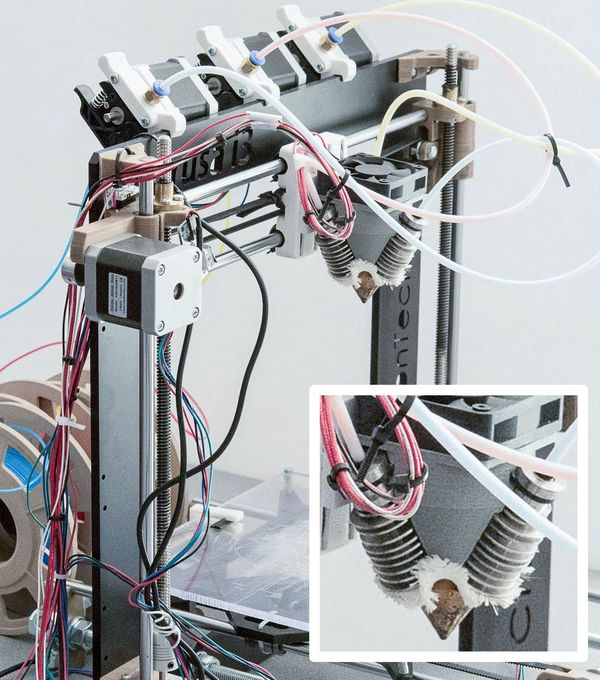}} \hfill
    \includegraphics[height=5cm]{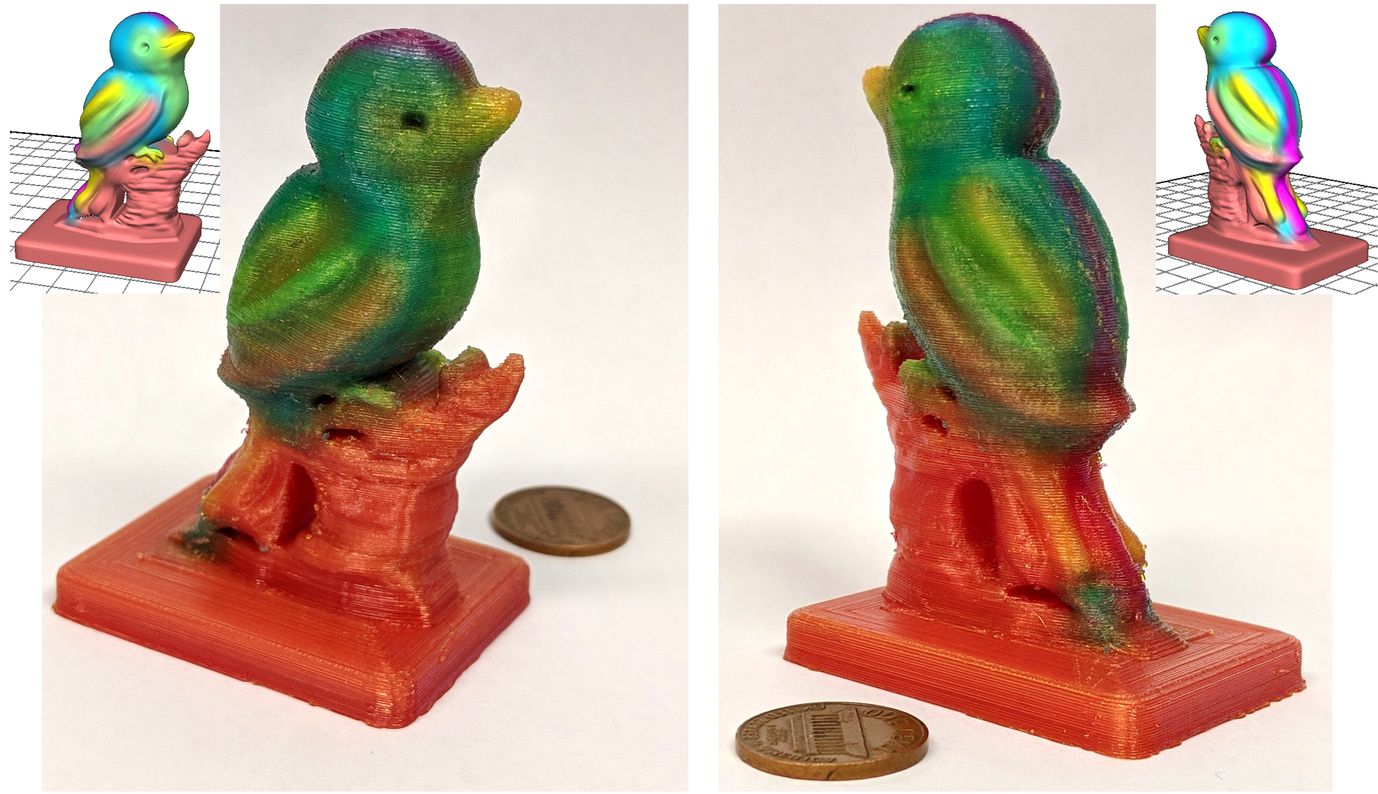} 
    \hfill
    \raisebox{0mm}[0pt][0pt]{%
    	\includegraphics[height=5cm]{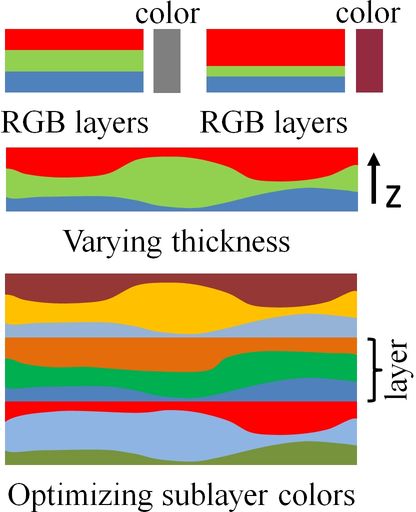}} \hfill
    \caption{
    \revised{
    Our technique enables color printing with precisely controlled gradients using a filament printer.
    \textbf{Left:} The printer is equipped with an off-the-shelf nozzle that inputs multiple filaments (three in the picture) and outputs molten mixed plastic through a single exit hole. Such devices cannot directly be used to print gradients: the transition between mixtures takes time, and this timing varies unpredictably.
    \textbf{Middle:} Using our technique, a user can 3D paint a virtual model and print an accurate, colored physical realization of her design. Printing uses standard, widely available materials (e.g. PLA, PET).
    \textbf{Right:} Our approach divides each layer into strata of varying heights to reproduce colors, printing them one on top of another.
	The color of each stratum is optimized to reduce visible stripes and print time.
	}
    }
    \label{fig:teaser}
\end{teaserfigure}

\maketitle

\section{Introduction}

Filament 3D printers have been widely adopted by hobbyists, enthusiasts, FabLabs, schools and small companies alike in the past few years. These printers fabricate a physical object from its numerical counterpart by accumulating melted material in successive layers: a filament is pushed through a hot nozzle where it melts and exits by a small hole. The molten material fuses with the layer just below and solidifies quickly as it cools down. By moving the extruder in the XY plane -- the build direction being along Z -- a solid layer of material is constructed. This principle can be implemented with widely available, inexpensive parts and electronics, and the RepRap movement produced many open source 3D printer designs. This, in turn, triggered a vast community effort to push the boundaries of fused filament fabrication: better software, new materials with specific properties (e.g. colored, transparent, conductive, ferromagnetic), and novel applications. Thanks to a constant improvement in software and modeling tools, the quality of printed parts has steadily increased.

While often regarded as a low-quality process, fused filament fabrication has several unique advantages. From the user perspective, it is inexpensive to operate (by a large margin) and involves less cleaning and precautions than resin and powder systems. Fabricated parts are also functional, as they retain many of the desirable properties of (e.g.) plastic materials.
From the fabrication perspective, the process offers unique degrees of freedom: most notably the ability to use multiple filaments and materials within the same print and the ability to freely move up and down the nozzle during deposition -- creating heightfield-like paths~\cite{Song:2017}.

Among the features that are expected from a 3D printer, and that are of special interest to Computer Graphics, is the ability to produce colorful prints. While this has been achieved for a few other technologies -- for which it nevertheless remains a topic of active research \cite{Babaei:2017:SIG,Brunton:2018:PSV} -- this ability is still lacking for filament printers, despite several attempts (see~\Cref{sec:prevwork}). This is especially unfortunate since many users are hobbyists printing decorative objects, toys, parts for model cars and planes, and boardgame figures for which colors and aesthetics are important. \revised{Color is also an effective way to carry information in the context of scientific visualization and education}.

\begin{figure}[tb]\centering
	\begin{center}
		\begin{overpic}[width=0.45\textwidth,trim=0 270 0 0,clip]{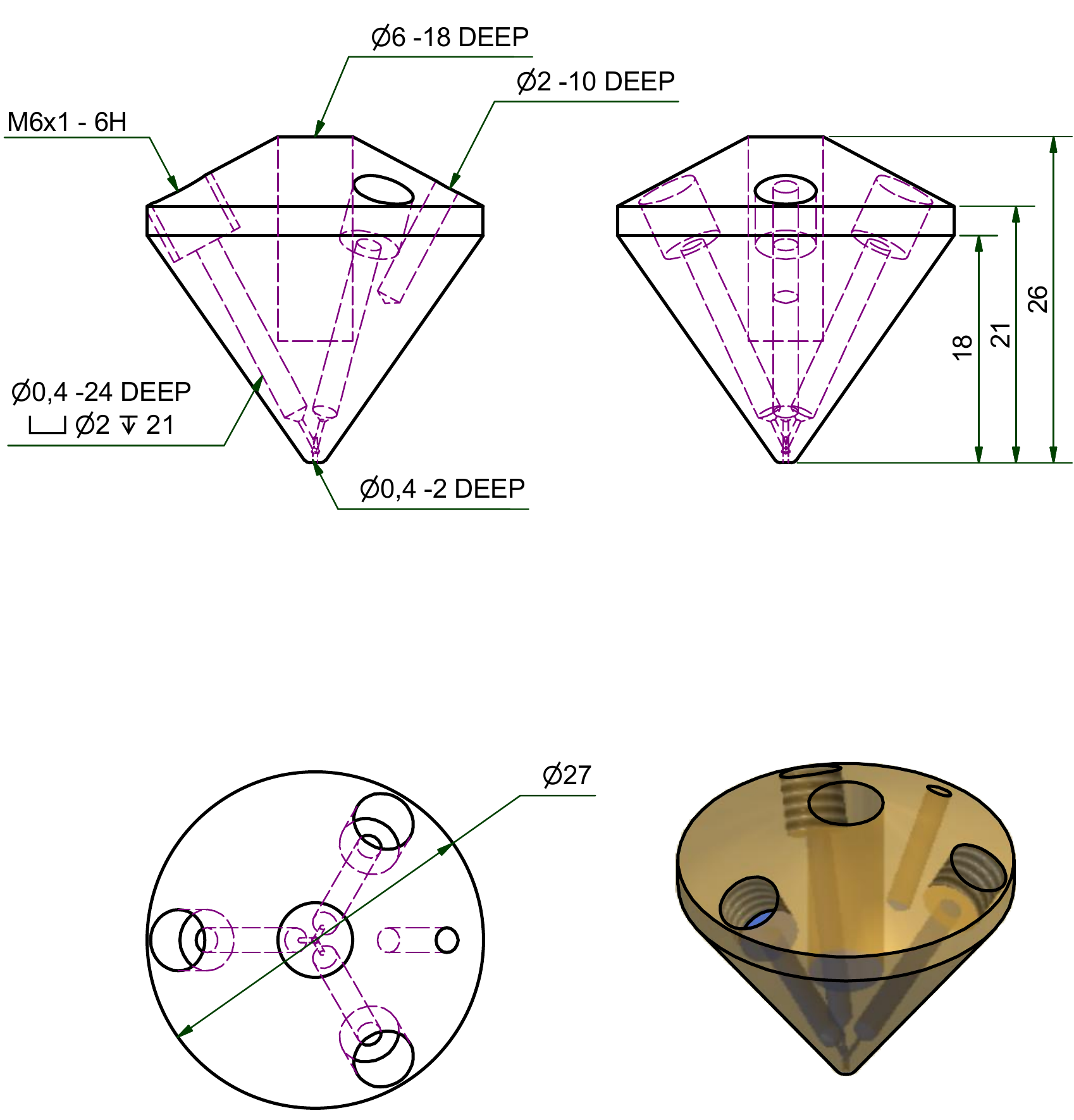}
		\end{overpic}
		\caption{Diamond nozzle design, blueprint by \cite{diamondnozzle}. Three filaments are pushed into the nozzle which has a single exit hole. A five filaments version is also available.}
		\label{fig:diamond_nozzle}
	\end{center}
\end{figure}

Our work proposes a novel technique that prints colorful parts reliably from multiple filaments. We exploit \textit{multi-in-one-out} nozzles that input multiple filaments and melt them in a unique chamber. These nozzles are inexpensive and available from several vendors (e.g. Diamond extruder~\shortcite{diamondnozzle}, Figure~\ref{fig:diamond_nozzle}).
It would seem that such a device is enough to provide color printing: one could expect to produce color gradients by mixing various ratios of colored filaments. Unfortunately, while this works for \textit{constant} ratios, dynamically changing the mix requires a significant transition time and cannot be controlled in a way that would produce reliable, precise {gradients} of colors (see \Cref{sec:colorfab:fdm}).

\revised{
	Instead, our method achieves colored prints by dividing each colored layer into a number of \textit{strata}. Each \textit{stratum} has a constant color, but this color is different from one stratum to the next (e.g. think of each stratum as being red, green, or blue). By slightly curving each stratum \textit{within} the layer thickness, adapting the plastic flow, the ratio of each filament in the final result is precisely controlled. This produces gradients of colors. The idea is illustrated in Figure~\ref{fig:teaser}, right. Note that the initial layers remain unchanged, only the strata within are optimized.
}

Our approach is facilitated by the multi-in-one-out nozzle in two ways. 
\revised{First and foremost, the ability to mix reliably at \textit{constant} ratios let us choose the color of each stratum. This allows to minimize defects such as visible stripe patterns and reduce print time as well as using much fewer strata than there are input filaments. To illustrate this, consider a constant color layer which is a mix of all filaments. Instead of using one stratum per filament, we can use a single stratum while mixing the filaments in the nozzle directly. Our approach generalizes this principle.}
Second, as there is a single nozzle mounted on the carriage, there are no calibration defects when depositing a different color for each stratum. This would be the case if switching between different nozzles mounted side by side~\cite{reiner2014dual,Hergel:2014:CCI,Kuipers2017}.

\vspace{1mm}
\noindent \textbf{Our contributions are:}
\begin{itemize}
	\item A novel approach to produce reliable color gradients in a 3D printed part, following an input color field, by depositing successive strata having each a constant color.
	\item An optimization procedure to choose the colors of strata such as to reduce color reproduction artifacts, and to allow the use of an arbitrary number of input filaments.
	\item A complete implementation running on inexpensive, widely available filament printer hardware.
\end{itemize}
Our technique is implemented in a standard slicer software.
It supports a varying number of input filaments, making it compatible with existing multi-in-one-out nozzles that input from two (e.g. \textit{E3D Cyclop}) to five input filaments (e.g. \textit{Full Color Diamond extruder}).

~\\ \noindent \textbf{Limitations. }
\revised{
	In this work we assume that the input specifies \textit{ratios} of the base filaments. The ratios are specified by the user, from e.g. a 3D printed palette, where each color corresponds to known mixing ratios (see Figure~\ref{fig:fish}). We do \textit{not} consider the color separation problem -- how to best approximate an arbitrary color field using the \textit{available} filaments -- this is left for future work. 
	We concentrate on the lower level challenge of depositing specified input ratios accurately while minimizing print time and defects.
}

\section{Related work}
\label{sec:prevwork}

The ability to print with multiple colors depends on the technology and is often linked to the ability
to print with multiple materials. We focus on color printing, with a particular
emphasis on filament printers, since that is our target. We only briefly mention other technologies
and multi-material printing.
We refer the interested reader to recent survey papers for more details \cite{Livesu:2017:cgf}.

\subsection{Color printing in additive manufacturing}
\label{sec:colorfab:others}

The main technologies capable of color printing apply colors on entire layers before solidification, e.g. inkjet on powder~\cite{zcorp} and inkjet on paper~\cite{mcor} for laminated fabrication.
DLP resin printers, which solidify layers by selectively exposing them to light, can perform multi-material printing through the use of multiple resin tanks \cite{Zhou:2011:DMM}.
Multi-jet technologies~\cite{polyjet,Sitthi-Amorn:2015:SIG}, which locally deposit and solidify droplets of different resins, also enable color printing.
The resins are tinted and tiny droplets of e.g. cyan, magenta, yellow are deposited following a dithering procedure.
Most processes only use a discrete number of base materials. Several approaches~\cite{Wu:2000:MFD,Cho:2001:MFD} discuss process planning to convert continuously varying material information into a limited set of base materials by a half-toning technique. 
\revised{
	Cignoni et al.~\shortcite{CGPS08} exploit color printing to improve the visual perception of geometric details of a printed object, taking into account sub-surface scattering.
	Brunton et al.~\shortcite{Brunton:2015:SIG} rely on half-toning and translucency in the context of color reproduction -- producing highly realistic colored 3D objects, including translucent appearances~\shortcite{Brunton:2018:PSV}.
	Elek et al. \shortcite{elek17scattering} introduce a technique to counteract heterogeneous scattering and obtain sharp details along 3D prints.
	Auzinger et al.~\shortcite{Auzinger:2018:CDN} explore how to manufacture structural colors at the nanoscale, without requiring pigments.
}
Babaei et al.~\shortcite{Babaei:2017:SIG} propose to layer colored resin droplets in small stacks normal to the surface. By varying the thickness of each colored layer within the stack, precise color reproduction is achieved. This exploits translucency through individual stacks rather than traditional dithering.

The major challenge we face is that filament deposition requires continuity when extruding filament, while the aforementioned high-end technologies can address (solidify) voxels individually, switching materials instantaneously.

\subsection{Color printing with filaments}
\label{sec:colorfab:fdm}

There is a strong interest in filament color printing. This is witnessed in particular in the hobbyist community, with several successfully funded community projects \cite{prometheus,nix,diamondnozzle,prusai3,rova4d}.
The main idea to achieve color printing with filament is to somehow blend filaments of different colors. This can be achieved in several ways.

\revised{
	A first approach is to use a printer having multiple hotends, typically mounted side by side on the same carriage, and loaded with filaments of different colors. 
	Colors are deposited alongside another in each layer.  This is made challenging by calibration issues and filament oozing. 
	Hergel and Lefebvre~\shortcite{Hergel:2014:CCI} specifically target these issues through path planning and part orientation.
}

\revised{
	Another approach is to use a single nozzle and switch filaments during printing. 
	This can be done manually\footnote{For impressive results of manual switching see
		\url{https://www.thingiverse.com/thing:25612}} with limited complexity (as this requires constant supervision/intervention). 
	The \textit{Palette} project \cite{palette} automates the process with a device that splices and joins multiple filaments into a single, continuous, multi-color filament. 
	Another automated approach is to rely on a \textit{switching extruder} \cite{prometheus,Builder,prusai3}, where a mechanical system allows for quick selection between different filaments.
	Using these devices reveals an interesting effect. There is always a transition between filaments: when a different material is pushed into the melting chamber, it mixes with the previous one. During this transition, the output material takes an intermediate color.
	To hide this transition, a \textit{purge tower} is printed alongside the print and disposed of afterward.
}

\begin{figure}[htp]
	\centering
	\begin{subfigure}[b]{0.48\linewidth}
		\centering
		\includegraphics[width=\textwidth]{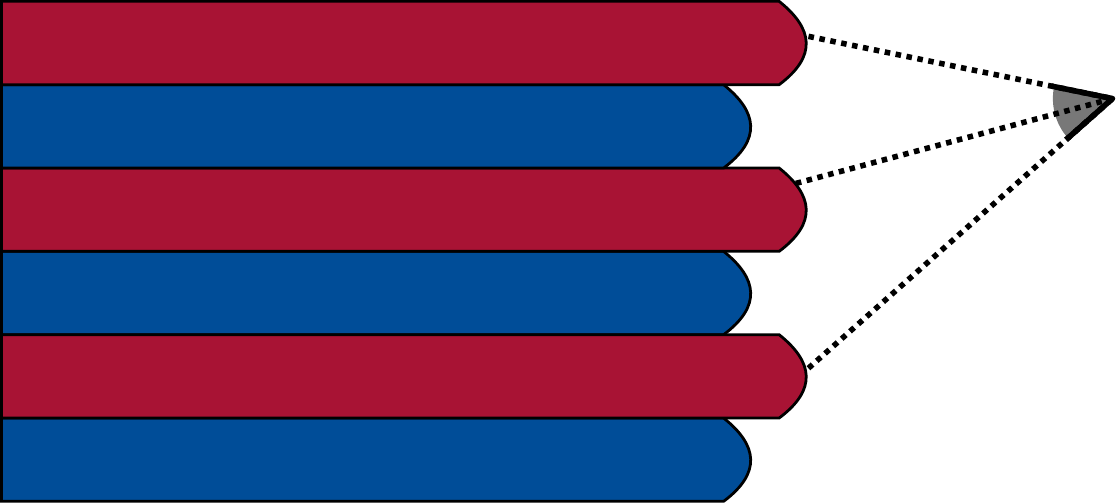}
		\caption{}\label{fig:sublayers_comparison:hatching}
	\end{subfigure}
	\hfill
	\begin{subfigure}[b]{0.48\linewidth}
		\centering
		\includegraphics[width=\textwidth]{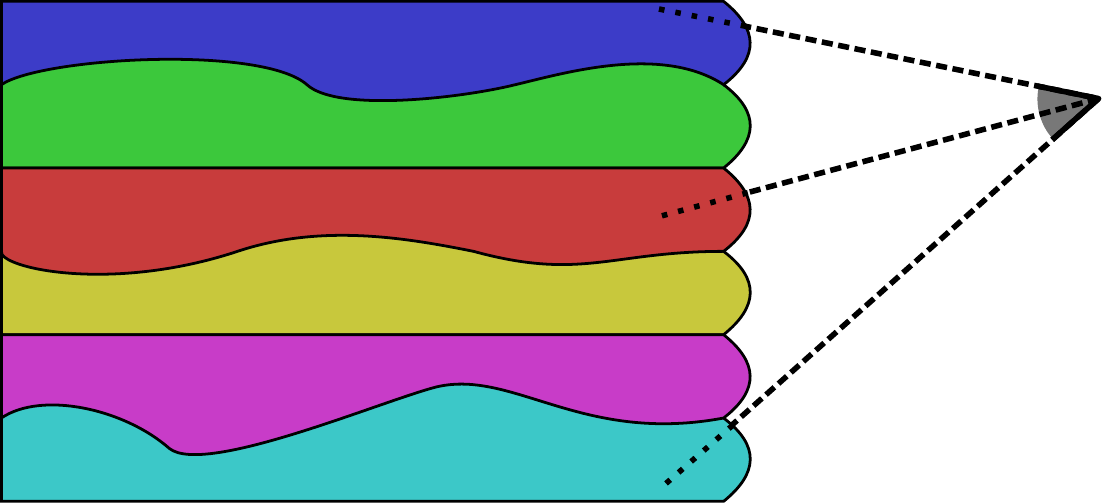}
		\caption{}\label{fig:sublayers_comparison:our}
	\end{subfigure}
	\caption{\revised{
			\textbf{(a)} Existing methods rely on dithering along the object surface. This is achieved by having layers
			protrude more or less from the side, changing the perceived color ratio. 		
			\textbf{(b)} In contrast our method 
			divides each layer into a number of strata, each having a different base color. By changing the thickness of each
			stratum, different ratios of the base colors are perceived. 
	}}
	\label{fig:vsdithering}
\end{figure}

\revised{
	The aforementioned methods select one filament at a time, and are not designed to produce mixtures. To create the perception of gradients, \cite{reiner2014dual} interweave filaments of different colors along the outer perimeter of a print. By changing which filament (color) is the most visible, a gradient effect similar to dithering is achieved. This approach, implemented in Photoshop CC \cite{photoshop}, can produce impressive results. However, the surface is covered by a high-frequency weaving pattern (sine wave), and results exhibit color deviations along slanted surfaces. A similar approach is proposed by \textit{Voxelizer} \cite{voxelizer}. Kuiper et al.~\shortcite{Kuipers2017} revisited this idea, generating dithering by pushing more or less flow on layers, which makes them protrude more or less on side surfaces (see Figure~\ref{fig:sublayers_comparison:hatching}). 
	Dithering has to be performed differently depending on the surface slope (vertical/horizontal). 
	With both methods, the perceived color is slightly impacted by the view angle, as occlusions play an important (see Section 5.4 of~\cite{Kuipers2017}).
	Instead of dithering, we rely on translucent filaments, depositing them in thin, slightly curved strata (see Figure~\ref{fig:sublayers_comparison:our}). This limits view dependency, does not require a slope-dependent algorithm and the effect extends beyond the surface, within the object volume.
}

\begin{figure}\centering
	\begin{center}
		\includegraphics[width=0.9\linewidth]{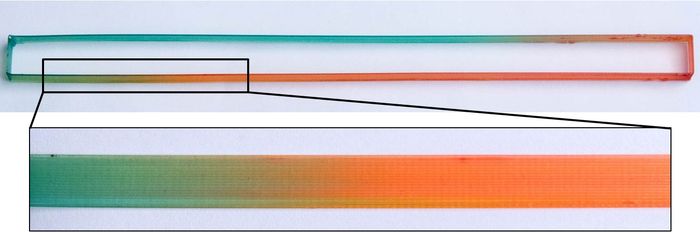}
		\caption{This vertically extruded rectangle ($170\times10$mm) is printed with the same orientation and exact same parameters at all layers, in a single continuous deposition path. The mix of filaments being pushed in the extruder is modified exactly halfway, switching from orange to green, and then from green to orange on the other side. It takes at least $57$ mm for the deposited color to change to the target color. Also, note how the gradients obtained at every layer vary unpredictably.}\label{fig:experiment_transition}
		\label{fig:badgrad2}
	\end{center}
\end{figure}

\revised{
	A final approach for color mixing is to rely on multi-in-one-out nozzles~\cite{richrap,rova4d,diamondnozzle,nix,Corbett:2012}, which allow to explicitly mix filaments. Filaments are pushed in the melting chamber according to a ratio. 
	%
	%
	While this sounds very promising to create gradients, there is a catch: the transition between different mixtures takes a non-negligible time. Our tests revealed that while we can easily determine after which volume a stable mix is achieved, the transition time is not stable and repeatable. Therefore, these extruders cannot produce a controlled gradient, and the results remain unsatisfactory. This is clearly visible in prints using mixing extruders, as shown in Figure~\ref{fig:badgrad2}.
	In addition, even if the transition time was precise, it remains too long compared to what would be required to reproduce spatially varying details at the scale of a 3D printed object.
}



\revised{
	Finally, a hybrid approach consists in tinting the filament during extrusion. A low-cost approach is to use colored markers~\cite{rainbow}. Company XYZ made public a prototype of a printer combining extruded filament with standard inkjet heads to produce colored 3D prints~\cite{inkfdm}. A special filament is required to enable the absorption of colored inks, while we seek to enable colored printing using a wide variety of available filaments as input (e.g. PLA, PET).
}

\section{Colored Fused Filament Fabrication}
\label{sec:color_mixing_theory}

Our approach is inspired by the way computer screens produce images. Each pixel physically corresponds to three tiny light emitting elements (red, green, blue). By varying the amount of light emitted by each, a different color is perceived.
Similarly, we decompose each layer into smaller strata -- a row of 'pixels' -- and vary the contribution of each stratum in the final result by changing its thickness, hence producing different colors.
However, contrary to a regular screen, and thanks to the use of a multi-in-one-out nozzle, we may choose the base colors of each stratum.
We exploit this additional degree of freedom to improve print quality and reduce print time.

\revised{
	While our technique is geared toward producing colored prints, we do not tackle the color separation problem (how to find the mixing ratios to best approximate a color~\cite{Babaei:2017:SIG}). Thus, we assume the user specifies colors as mixing ratios of the base
	filament mounted on the printer (e.g. picking them from a 3D printed palette, see Figure~\ref{fig:fish}).
	Our goal is to provide the low-level technology that will enable full-color printing in the future. 
	However, more work will be required to elaborate filaments and to model which colors are perceived for a given mixture. We leave these
	challenges as future work. Thus, in the remainder of the document, we will no longer refer to \textit{colors}, but instead
	to \textit{mixing ratios}.
}

\subsection{Input}
\label{sec:input}

The input to our approach is a set of slices to be printed (toolpaths), as well as a volume field specifying mixing ratios.
The toolpaths are obtained from a standard slicer, which computes deposition paths after slicing the input object into a set of layers.
The volume field is a function which returns the mixing ratios producing the desired color at any point in space.
This function may be a procedure, an access into a volume texture, or any other similar mechanism.
In our implementation, we use a volume texture enclosing the object and let the user paint the model with a 3D brush.

When using $K$ base filaments ($K>1$) we denote a mixing ratio as a vector of $K$ weights $c \in \mathbb{R}^K$ such that $\sum_{i=1}^{K}{c_i} = 1$ and $c_i \geq 0$.
This ensures the input ratios can be physically reproduced by
mixing the base filaments on the target printer.
\revised{Note that this is a requirement: we do not support converting mixing ratios for e.g. a five 
	filament printer into mixing ratios for a printer having fewer filaments.}



\subsection{Output}

The output of our technique is a G-code file for a printer equipped with a multi-in-one-out nozzle.
For the sake of clarity, let us assume a nozzle with three input filaments (we use both three and five filament printers). 
Extrusion is controlled through modified G-code commands that allow specifying the ratios of each filament in the mix, e.g. G-code command \texttt{G1 X10.0 Y12.0 Z3.0 E20.5 A0.2 B0.3 C0.5} moves the nozzle to position (x=$10.0$mm, y=$12.0$mm, z=$3.0$mm) and extrudes filament up to length $20.5$ mm, using ratios of $20\%$, $30\%$ and $50\%$ for the filaments in extruders A, B and C. 
The ratios have to be positive and have to sum to one.

The strata are slightly curved to produce the mixing ratio variations. Thus, along a toolpath we vary the z coordinate by small amounts, and we adjust the flow accordingly -- by changing motion speed -- to maintain a constant track width. Each stratum is a sequence of G-code instructions with varying Z coordinates and E values (amount of material), where the variations are given by the mixing ratios and the shape of the stratum below, see Figure~\ref{fig:teaser}, right.
The strata, taken together, reproduce the shape of the original layer.


\subsection{A first approach and its limitations}
\label{sec:motivation}

Given the idea of using strata, a first approach would be as follows.
Each base color would be mapped to one stratum, following an arbitrary fixed order (e.g. red first, then blue, then green).
Then, strata would be printed in sequence one on top of another, following the geometry of the original toolpath, but with slight height variations.
This ideally creates curved strata that produce the desired effect.

Unfortunately, this simple approach does not produce satisfactory results and suffers from three issues: \textit{stripes},  \textit{interferences}, and \textit{non-scalability}. 

\begin{figure}[tb]\centering
	\begin{center}
		\includegraphics[width=0.6\linewidth]{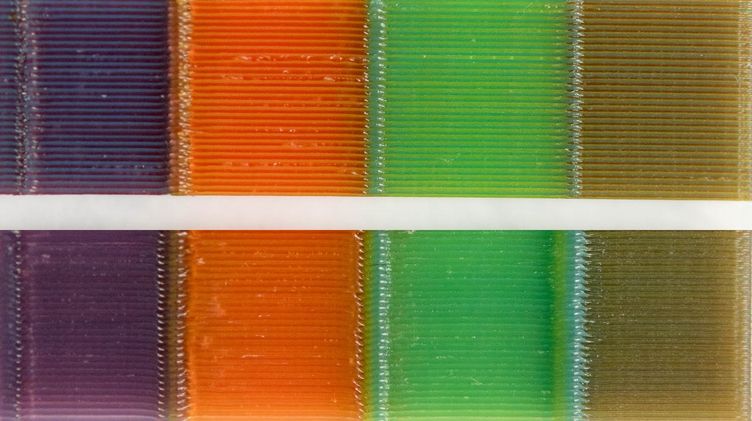}
	\end{center}
	\caption{\textbf{Top:} Using pure filaments as base mixtures produces visible stripes. 
		\textbf{Bottom:} A better choice of strata mixtures makes stripes barely noticeable. }
	\label{fig:stripes_basic}
\end{figure}

\noindent \textbf{Stripes} (Figure~\ref{fig:stripes_basic}). The typical layer thickness we use is $300 \mu m$. Thus, the strata are relatively thick, and the stripe pattern produced by interleaving the base colors may become visible in the final result. This is especially the case when mixing two colors with large ratios (e.g. a 50\% blend of two base colors).

\noindent \revised{ \textbf{Interferences} (Figure~\ref{fig:fringes_printed}). Undesired colors may appear between layers. This is akin to the color fringing artifacts around white text on computer screens. The reason is that the color is perceived as the average of surrounding strata, regardless of layer boundaries. Thus, the last stratum of layer $i$ will perceptually mix with the first stratum of layer $i+1$, possibly resulting in a spurious color. 
}

\noindent \revised{\textbf{Non-scalability}. Finally, the simple approach described here always requires one stratum per base filament. Thus, each additional source filament requires an additional stratum: the method does not scale with the number of source filaments. 
	The number of stratum should be kept low. Each added stratum increases print time: a purge sequence is required to stabilize the new mixture in the nozzle, and then the stratum has to be printed. In addition, using many strata requires careful calibration as average deposition thickness decreases. We successfully print using five strata on $0.4$ mm layers (see Figure~\ref{fig:5d}), but going beyond seems unreasonable.
}

To address all three issues, we propose to optimize both the number of stratum per layer and their mixtures, using the ability of the mixing extruder to reliably produce \textit{constant} mixes of base filaments. 
\minor{This is based on the observation that the perceived color is similar whether the mixture is done within the nozzle or through strata of varying heights.
}
As most objects exhibit a strong spatial coherency within a layer, it is rare for a painted object to contain all possible colors~\footnote{This fact is at the root of many texture compression algorithms, such as DXTC and S3TC.}. This implies that we can optimize a choice of mixtures that will 1) minimize the stripes by using base mixtures closer to that used within the layer and 2) decouple the number of strata from the number of base filaments.

\subsection{Optimization of strata base mixtures}
\label{sec:optimization}

\revised{We now detail our optimization process. The objective is to compute the base mixture used for each stratum as well as reduce the number of strata, exploiting the mixing extruder as much as possible.
	Our formulation allows for arbitrary numbers of filaments, but in practice our most complex printer takes only five. 
	At most, using $K$ filaments would require printing $K$ strata per layer. Our optimization reduces this number automatically, when possible.}

\revised{We formulate the optimization in two steps. 
	First, we optimize the number of strata and their base mixing ratios (the ones used with the nozzle, when printing each stratum). 
	Second, we compute an ordering of the strata within each layer to obtain the final result.}

\revised{The optimizer does not change the quantity of each filament deposited at any given location in the print:
	it simply makes a compromise between mixing in the nozzle or mixing using curved strata.}

\subsubsection{Optimizing number of strata and their base mixtures}

\revised{
	This optimization step is performed independently for each layer. We, therefore, consider a single layer in the following.
}

\revised{
	Before processing, we re-sample the layer toolpaths: we increase the number of vertices to properly capture variations of the input field. 
	The sampling ratio is constant.
}

\revised{
	We denote the mixing ratio associated with each toolpath vertex by $c \in \mathbb{R}^K$ such that $\sum_{i=1}^{K}{c_i} = 1$ and $c_i \geq 0$.
	Note that they are akin to barycentric coordinates.
}

\begin{figure}[htp]
	\centering
	\includegraphics[width=0.6\linewidth]{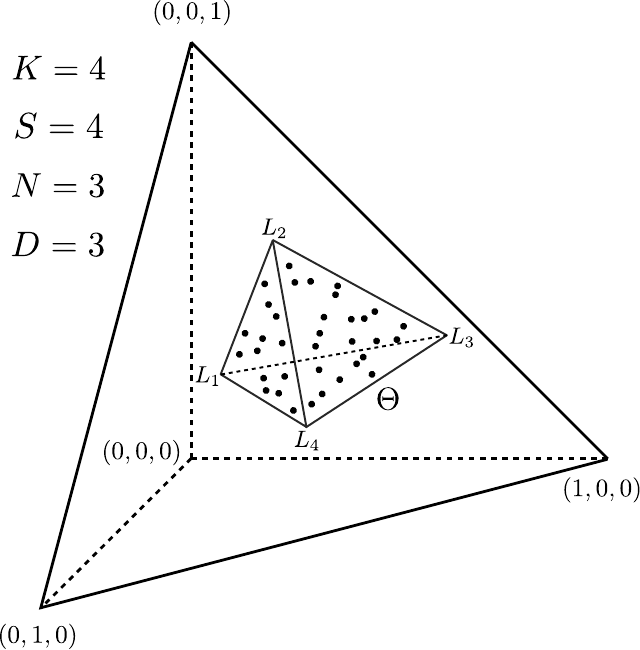}
	\caption{
		\revised{Four filaments case. The mixing ratios are embedded into 3D using unit vectors and the origin, forming the set $\mathcal{C}$ (black dots).
			Our optimizer finds a simplex $\Theta$ (here tetrahedra) enclosing $\mathcal{C}$ while having a small volume. The vertices
			of the simplex ($L_1,L_2,L_3,L_4$) are the new strata base mixtures. Since they are closer to the enclosed mixing
			ratios, the stripe effect is reduced.}
	}
	\label{fig:optimization_general}
\end{figure}

\begin{figure}[htp]
	\centering
	\begin{subfigure}[b]{0.60\linewidth}
		\centering
		\includegraphics[width=\textwidth]{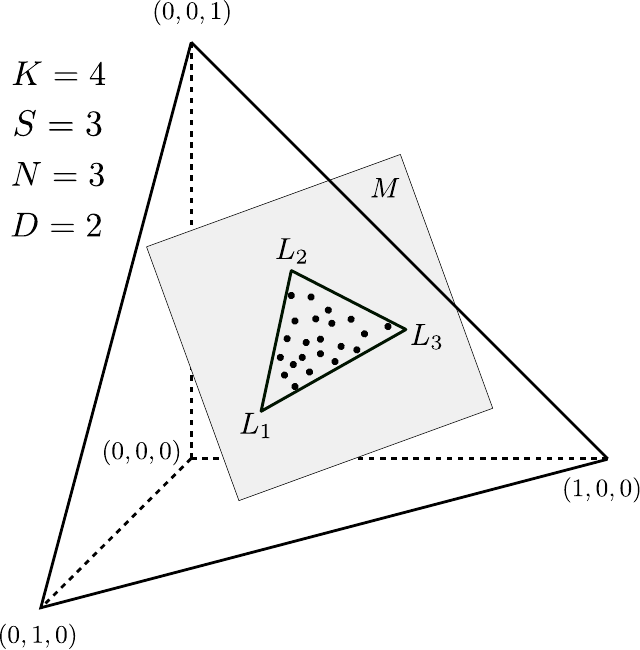}
		\caption{}\label{fig:optimization:space}
	\end{subfigure}
	\hfill
	\begin{subfigure}[b]{0.35\linewidth}
		\centering
		\raisebox{1cm}{\includegraphics[width=\textwidth]{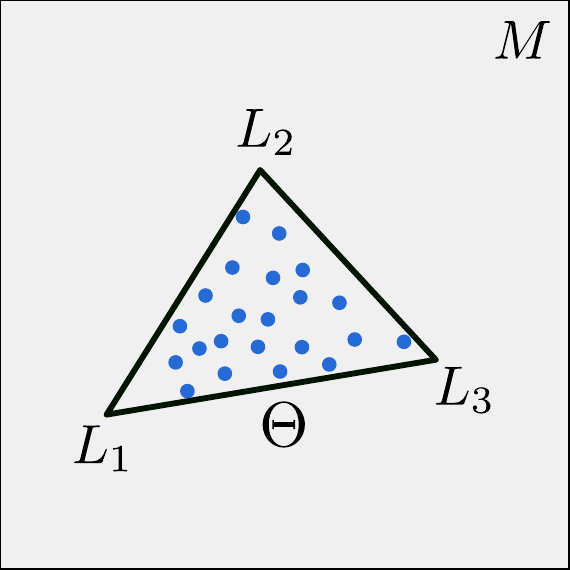}}
		\caption{}\label{fig:optimization:projection}
	\end{subfigure}
	\caption{
		\revised{Four filaments case. (\ref{fig:optimization:space}) Here, the mixing ratios $\mathcal{C}$ lie on a lower dimensional subspace.
			(\ref{fig:optimization:projection}) Our algorithm first determines the intrinsic dimensionality of the data (2D) and
			computes a dimensionality reduction to obtain the set $M \otimes \mathcal{C}$ (in blue).
			It then optimizes for a simplex in the lower dimension space (here, a triangle).
			Only three strata will be used in this case to accurately reproduce the input mixtures.
		}
	}
	\label{fig:optimization_reduction}
\end{figure}

\revised{
	Our objective is to determine the number of required strata $S$ for the layer ($1\leq S\leq K$),
	as well as the base mixtures of each stratum. We denote each of the $S$ base mixture by $L^i \in \mathbb{R}^K$, with $1 \leq i \leq S$.
	These mixtures follow barycentric constraints: $\sum_{j=1}^K{L^i_j} = 1$ and $L^i_j \geq 0$.
	Ideally, we would like the $L^i$ to be close to the input ratios, such that the stripe effect
	is minimized.
}

\revised{
	After optimization, each input mixing ratio $c$ will be converted into a new mixing ratio $\alpha$ (also under
	barycentric constraints). The $L^i$ are used as mixing ratios with the nozzle, while the $\alpha$ are used
	to change the thicknesses of the strata during printing.
}

\revised{
	A key observation, implied by the barycentric constraints, is that all the input ratios $c$ lie 
	within a simplex defined by the $L_i$.
	Therefore, optimizing for the $L_i$ amounts to finding a simplex $\Theta$ enclosing all input points,
	and such that its vertices follow barycentric constraints. 
	This is illustrated in Figure~\ref{fig:optimization_general}.
	To minimize the stripe defect we minimize the volume of $\Theta$, which ensures the $L_i$ are close to the
	points they enclose.
}

\revised{
	Since the mixing ratios represent barycentric coordinates, we embed them into a space of dimension $N = K-1$ ($N > 0$) using
	the origin and unit vectors. The loss of one dimension is expected, as for instance when using two filaments 
	a single number is enough to represent the mixture.
	We denote by $\mathcal{C} \subset \mathbb{R}^N$ the set of all embedded mixing ratios for the layer. 
}

\revised{
	The number of required strata is directly impacted by the intrinsic dimensionality of $\mathcal{C}$. For instance,
	if all points lie on a line, only two strata will suffice regardless of the value of $K$: all points will be linear
	combination of two base mixtures.
	Therefore, if possible, we seek to compute a dimensionality reduction of $\mathcal{C}$. 
	Let us denote by $M$ and $M^{-1}$ the dimensionality reduction operator and its inverse, 
	as well as $D \leq N$ the intrinsic dimensionality of the data.
}

\revised{
	In dimension $D$, we search for the smallest simplex $\Theta$ enclosing the dimensionally 
	reduced points $M \otimes \mathcal{C}$, ensuring the simplex vertices mapped back through $M^{-1}$ (the $L_i$)
	follow barycentric constraints.
	This is illustrated in Figure~\ref{fig:optimization_reduction}.
}\\

\noindent \textbf{Solver}. \revised{For dimensionality reduction ($M$ and $M^{-1}$), we rely on Principal Component Analysis (PCA).
	This provides us with a transformation (defined by the normalized eigen vectors) aligning the first coordinates 
	with the dimensions having the largest variance. We select the number of dimensions using a threshold $\epsilon$ 
	on the eigen values (variance). We set $\epsilon = 10^{-4}$ -- it is defined in ratio space -- to keep
	the approximation error low.
}

\revised{
	Minimizing the volume of $\Theta$ corresponds to a known, hard combinatorial optimization problem~\cite{Hendrix:2013:MVS}.
	A known property of the optimal convex hull $\Theta$ is that each of its $N$-dimensional facets must touch the convex hull of $\mathcal{C}$.
	This property was used in~\cite{Zhou:2002:AMV} to obtain an algorithm enumerating all feasible contacts for $N=3$.
}

\revised{
	We follow a similar idea to propose an approximate algorithm that extends to any dimension.
	Our algorithm takes into account only one type of contact: we enumerate all combinations of $N + 1$ hyperplanes of the convex hull 
	of $\mathcal{C}$ that enclose a finite volume. We compute the volume of each valid combination -- that is those which 
	vertices enforce barycentric constraints through $M^{-1}$ -- and return the one with minimal volume.
	To help finding solutions we allow a small tolerance when checking for barycentric constraints, using a 
	threshold $\lambda = 10^{-2}$ on all checks (sum to one and positivity).
}
\revised{
	In case no solution is found, we increase $D$. At worst, we end with $D = N$. If still no solution is found, 
	we use the trivial solution of setting $L_i$ to be unit vectors.
}
\revised{
	This process is fast. On average, the number of convex hull hyperplanes grows slowly with respect to the number of points $\left|\mathcal{C}\right|$.
	For instance, for a random distribution of points inside a polytope, the average number of hyperplanes is $\Theta(\log^{N-1}{\left|\mathcal{C}\right|})$~\cite{Dwyer:1988:AAA}.
	Optionally, we could randomize the enumeration and limit the maximum amount of tests to improve efficiency.
	We use QHull~\cite{Barber:1996:QAC} for computing the initial convex hull in N-dimensions.
}


\subsubsection{Optimizing the ordering of the strata}

\begin{figure}[tb]
	\includegraphics[width=0.7\linewidth]{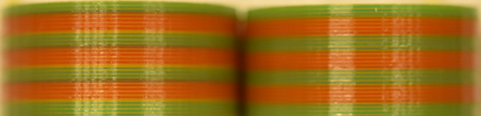}
	\caption{\textbf{Left:} Some ordering produce visible interferences between neighboring layers. \textbf{Right:} Reordering the strata removes these defects.}
	\label{fig:fringes_printed}
\end{figure}

\revised{
	After optimizing for the strata mixtures we need to determine their ordering
	in the final print. Some orderings are better than others: an ordering
	where strata with similar mixtures in successive layers end up side by side
	will likely produce a visible interference (see Figure~\ref{fig:fringes_printed}). 
	Instead, we encourage a sequence of as-different-as-possible strata,
	to obtain an interleaving pattern akin to LED screens subpixels (RGB,RGB,...).
	As the number of strata and their mixtures change with each layer, we rely
	on a combinatorial optimization process.
}

\revised{
	Our objective function evaluates whether similar mixtures, across layers, end up 
	close to each other in the final result. Since mixtures are used in varying amounts
	along and within the object, we take the mixture volume into account: 
	a mixture that is used only in a few places will have a limited influence on the ordering.
}

\begin{wrapfigure}[5]{r}{2.5cm}
	\vspace*{-0mm}
	\hspace*{-8mm}
	\includegraphics[width=3cm]{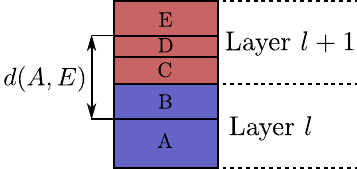}
\end{wrapfigure}
\revised{We denote $L^l_i$ the $i$-th base mixture of layer $l$, and denote by $d(A,B)$ 
	the \textit{stack distance} between two base mixtures. Considering an imaginary stack 
	of all the ordered base mixtures, where the volume is used as height, the stack 
	distance is defined as the distance between the top of A and the bottom of B (see inset).
	Then, a current ordering is scored as:
	\[
	\sum_{i,j}\left( \frac{\|L^{l}_j-L^{l+1}_i\|}{1 + d(L^{l}_j,L^{l+1}_i)} \right)
	\]
	where $i,j$ pairs are pairs of indices for respectively strata in layer $l+1$ and $l$.
	We seek an ordering with a maximal score.
	The principle is that pairs which have similar mixtures (numerator) produce a little score,
	while pairs with dissimilar mixtures produce an increasing score when further apart in
	the stack (denominator). The volume is taken into account through the stack distance.
}

\revised{
	Our approach is a simple bottom-up sweep, determining the ordering of layer $l$ assuming
	the ordering of layer $l-1$ is known. At the current layer $l$
	we enumerate all possible permutations of strata (there are five strata maximum in our
	implementation), and select the one maximizing the objective. The first layer
	is initialized with a random ordering.
}



\section{Implementation}
\label{sec:implementation}

We implemented our approach in a custom slicer.
After generating the toolpaths, we optimize for the per-layer strata, and generate new path planning instructions.
Table~\ref{tbl:times} gives processing times for our main results ; all complete in less than four seconds.


We produce the G-code by outputting instructions for each layer as many times as there are strata, adjusting the height and the flow along the paths.
The height at each vertex is directly obtained from the barycentric coordinates $\alpha$.
We then simplify the geometry of the obtained curved paths to reduce the number of vertices.
Note that when the height of a path vanishes, we interrupt deposition and proceed to the next path.

\revised{
	An important factor in achieving successful prints while using several strata and thin layers is to
	implement what is known as \textit{linear advance}\footnote{\url{http://marlinfw.org/docs/features/lin_advance.html}}.
	This is a mechanism to deal with internal pressure within the nozzle, that stops pushing plastic before the end of a path, 
	such that pressure is reduced when travel motions start.
}

Before each stratum, we print an auxiliary structure that allows the new filament mix to stabilize. 
We chose to use an ooze shield structure \cite{Cura,hornus:2016} as it remains close to the print.
We adjust print speed to maintain a constant extrusion speed, which avoids abrupt pressure changes in the nozzle~\cite{Kuipers2017}.



\section{Results}
\label{sec:experiments}


To evaluate our results, we built several custom printers using a Prusa i3 kit and a Diamond extruder \revised{(three and five filaments)}. One of the printers is shown in Figure~\ref{fig:teaser}, left.
We perform all computations on an Intel Core i7, CPU 4.0 GHz, with 32 GB memory.
Unless otherwise specified we use a layer thickness of $0.3$ mm.

\begin{figure}[tb]
	\includegraphics[width=0.6\linewidth]{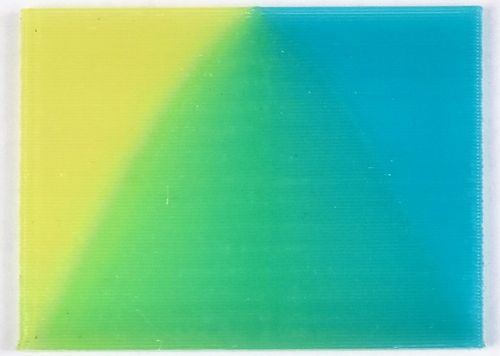}\\
	\includegraphics[width=0.99\linewidth]{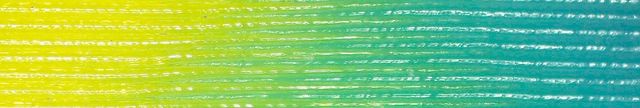}
	\caption{Gradient of varying sharpness, from smooth (bottom) to sharp (top). Our approach affords for a progressive transition. The bottom microscope close-up of the gradient tip reveals the interleaved strata.
		Please refer to the text for details regarding the small discontinuity on the left/right sides.}
	\label{fig:grad}
\end{figure}

~\\ \noindent \textbf{Gradients. }
%
Figure~\ref{fig:grad} shows a part with a gradient that starts smooth at the bottom and becomes progressively sharper until becoming an edge.
As can be seen, our method successfully copes with the transition. To the best of our knowledge, no other method can achieve similarly smooth and
progressive results on filament printers.

A small discontinuity can be seen just before transitioning to a pure color, on the left and right sides. 
This is due to deposition interruption when height vanishes, explained in Section~\ref{sec:implementation}. 
A very careful calibration of linear advance could reduce this effect further.

\begin{figure}[tb]\centering
	\begin{center}
		\hfill
		\begin{overpic}[height=2cm]{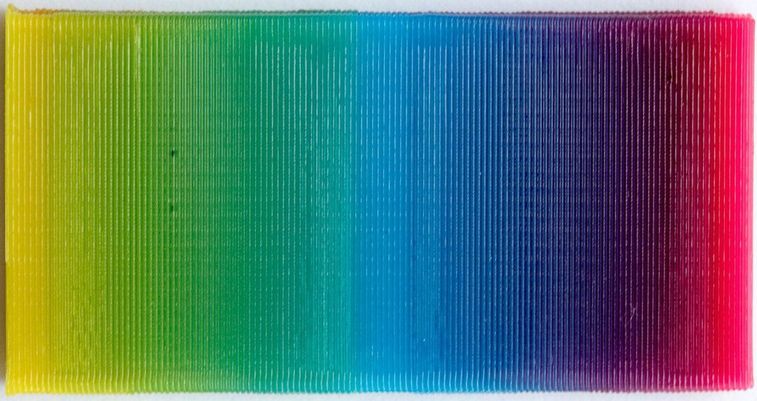}
		\end{overpic} \hfill
		\begin{overpic}[height=2cm]{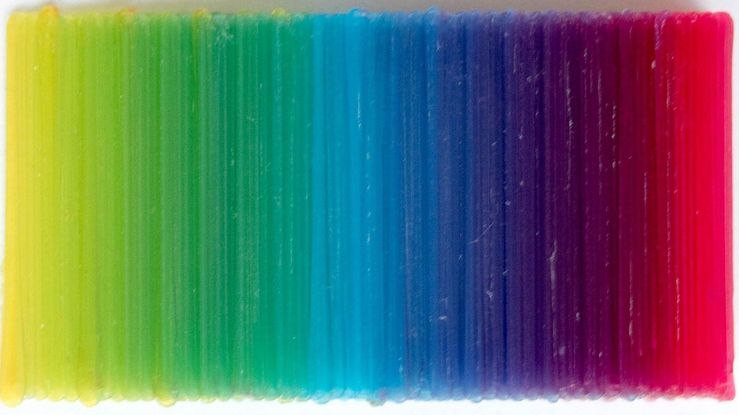}
		\end{overpic}
		\hfill
		\caption{Grading from yellow, cyan, to magenta.
			\textbf{Left:} Vertical print with gradient along the Z build direction.
			\textbf{Right:} Print with 27$^{\circ}$ slope, gradient along X. Thanks to translucency, the slope has little impact on the result.
		}
		\label{fig:slant_grading}
	\end{center}
\end{figure}

~\\ \noindent \textbf{Impact of the surface slope. }
The strata are printed one on top of another. While this works well along vertical surfaces, one concern is that on near-flat surfaces the last printed stratum might hide those below, biasing the perceived colors~\cite{Kuipers2017,reiner2014dual}. As mentioned in \Cref{sec:prevwork} we rely on slightly translucent filaments to avoid view dependency. A positive side effect is that this also alleviates the visibility issue on top layers, as can be see in Figure~\ref{fig:slant_grading}, right.

Using translucent filaments however introduces a drawback: a thicker colored shell has to be printed to hide the object interior, and the final result remains slightly translucent.
In our experience, most users actually find this effect pleasing and qualitative compared to opaque plastic prints.

\begin{figure}[tb]\centering
	\begin{center}
		\hfill
		\begin{overpic}[height=4cm]{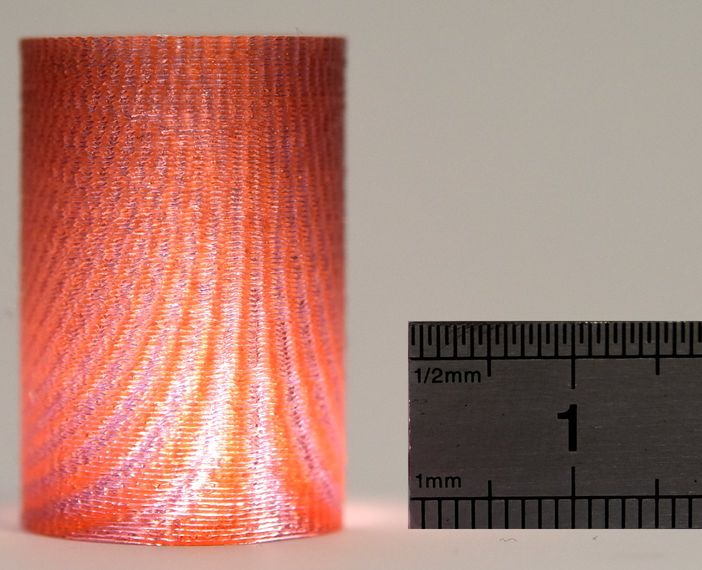}
		\end{overpic} \hfill
		\begin{overpic}[height=4cm]{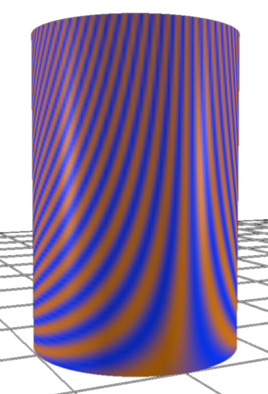}
		\end{overpic}
		\hfill
	\end{center}
	\caption{\revised{Testing achievable detail frequencies. This pattern is produced by a sine wave around the cylinder that has increasing frequency towards the top. The nozzle diameter is $0.4$ mm. The smallest features being properly captured are around $0.6$ mm. }}
	\label{fig:frequency}
\end{figure}

~\\ \noindent \textbf{Frequency. }
\revised{
	Figure~\ref{fig:frequency} shows a cylinder textured with a pattern of increasing frequency (a sine wave). As can be seen, high frequencies can be reproduced -- the main limitations stemming from the nozzle deposition diameter ($0.4$ mm in our setup). In this print, the thinnest visible stripes -- before aliasing occurs -- have a width of approximatively $0.6$ mm.
}

~\\ \noindent \textbf{Colored prints. }
We painted and 3D printed a variety of objects using our approach using PLA filament.
Unless otherwise specified prints use three source filaments and $0.3$ mm layers.

When considering the results, please keep in mind that there is no calibration between the colors displayed in our paint tool and the colors on the print: we faithfully reproduce the \textit{mixing ratios} selected by the user but have currently no way to display matching colors on the screen. 
In the results, however, regions of smooth gradients and sharp transitions should match accurately between the paint and the print.
\revised{The filaments used vary between prints, therefore the obtained colors differ.}

\begin{figure}[tb]
	\begin{center}
		\begin{subfigure}[b]{0.49\linewidth}
			\begin{overpic}[height=3.2cm]{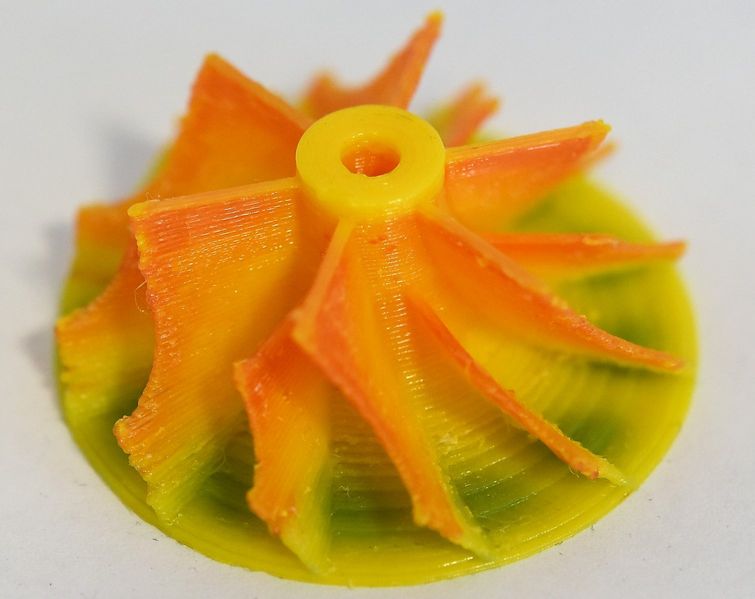}
			\end{overpic}
		\end{subfigure}
		\begin{subfigure}[b]{0.49\linewidth}
			\begin{overpic}[height=3.2cm]{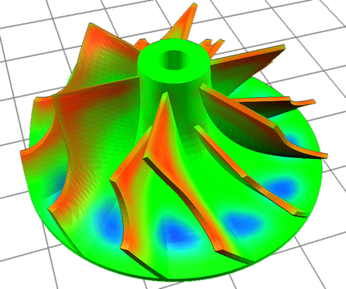}
			\end{overpic}
		\end{subfigure}
		\caption{\revised{3D printed turbine (Thing: 398841 by bob345). 
				\textbf{Left:} 3D painted model in our tool. 
				\textbf{Right:} photographs of the 3D printed model with a matching viewpoint.}}
		\label{fig:turbine}
	\end{center}
\end{figure}

The bird in Figure~\ref{fig:teaser} shows how a variety of colors can be used within the same print, from pure, clear colors to more subtle mixes. Gradients and sharp transitions are properly captured. Note how the top of the bird head remains properly colored despite the change in slope.
\revised{The turbine in Figure~\ref{fig:turbine} illustrates how our technique can be used for educational and prototyping/engineering purposes, visualizing a fluid pressure field directly on a 3D printed prototype.}

\begin{figure}[tb]\centering
	\begin{center}
		\begin{subfigure}[b]{0.49\linewidth}
			\begin{overpic}[width=0.9\textwidth]{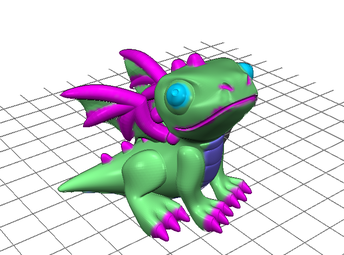}
			\end{overpic}
		\end{subfigure}
		\begin{subfigure}[b]{0.49\linewidth}
			\begin{overpic}[width=0.9\textwidth]{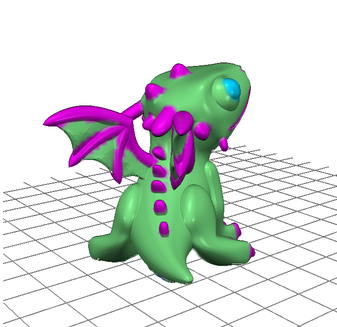}
			\end{overpic}
		\end{subfigure}
		
		\begin{subfigure}[b]{0.49\linewidth}
			\begin{overpic}[width=0.9\textwidth]{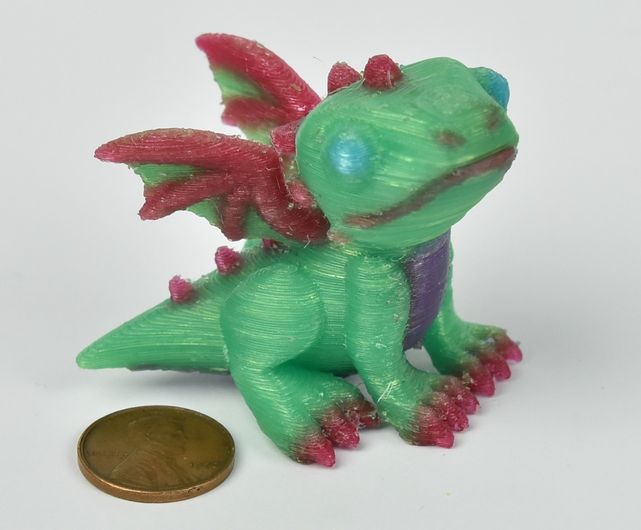}
			\end{overpic}
		\end{subfigure}
		\begin{subfigure}[b]{0.49\linewidth}
			\begin{overpic}[width=0.9\textwidth]{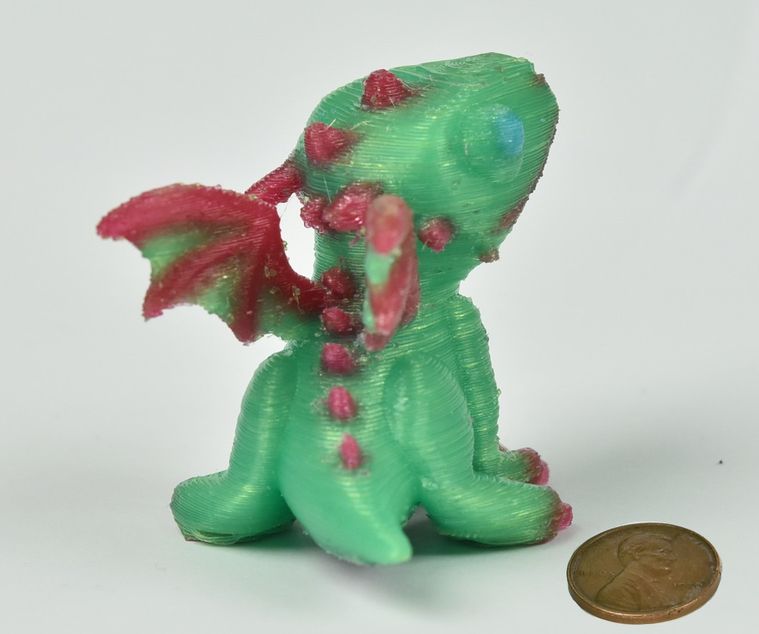}
			\end{overpic}
		\end{subfigure}
		\caption{3D printed dragon (Thing: 1624412). 
			\textbf{Top:} two views of the 3D painted model in our tool. 
			\textbf{Bottom:} photographs of the 3D printed model using our technique with matching viewpoints. } 
		\label{fig:dragon} 
	\end{center}
\end{figure}

\begin{figure}[tbh]\centering
	\begin{center}
		\begin{overpic}[width=0.99\linewidth]{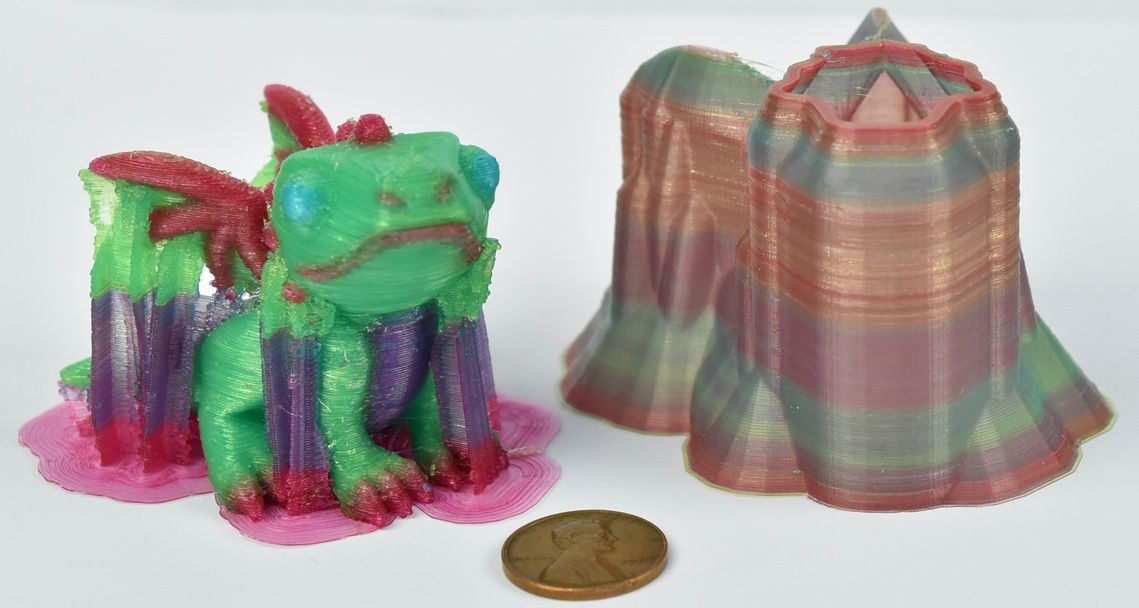}
		\end{overpic}
		\caption{\revised {3D printed dragon with its supports and its priming shield.}}
		\label{fig:shield}
	\end{center}
\end{figure}

The dragon in Figure~\ref{fig:dragon} reveals how printed colors closely follow the layout shown on the virtual model, with both gradients and sharp transitions faithfully reproduced.
Figure~\ref{fig:shield} shows the prime shield used for transitioning between strata mixes.

\begin{figure}[tbh]
	\centering
	\begin{center}
		\begin{overpic}[width=0.99\linewidth]{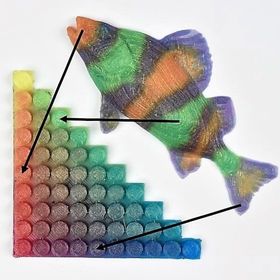}
		\end{overpic}
		\caption{3D printed fish and palette used to select colors (Thing: 1701871).}
		\label{fig:fish}
	\end{center}
\end{figure}

The fish in Figure~\ref{fig:fish} is a case of near-flat printing. As can be seen, there is no perceivable difference in quality between vertical and quasi-flat surfaces.

\begin{figure}[tbh]
	\centering
	\begin{center}
		\begin{subfigure}[b]{0.49\linewidth}
			\begin{overpic}[width=0.9\textwidth]{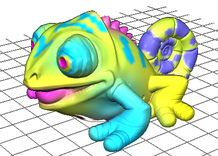}
			\end{overpic}
		\end{subfigure}
		\begin{subfigure}[b]{0.49\linewidth}
			\begin{overpic}[width=0.9\textwidth]{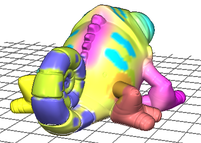}
			\end{overpic}
		\end{subfigure}
		
		\begin{subfigure}[b]{0.49\linewidth}
			\begin{overpic}[width=0.9\textwidth]{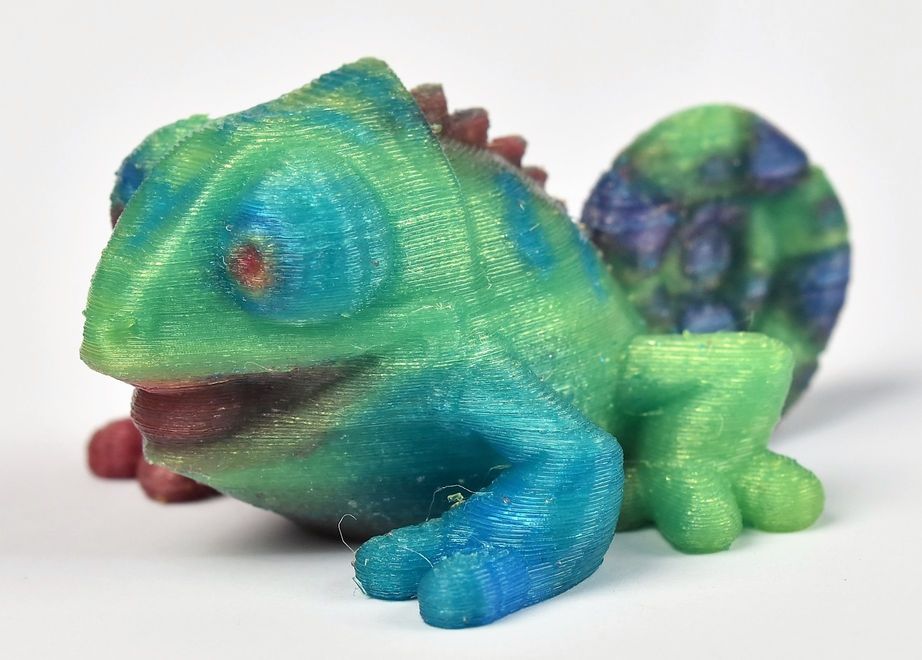}
			\end{overpic}
		\end{subfigure}
		\begin{subfigure}[b]{0.49\linewidth}
			\begin{overpic}[width=0.9\textwidth]{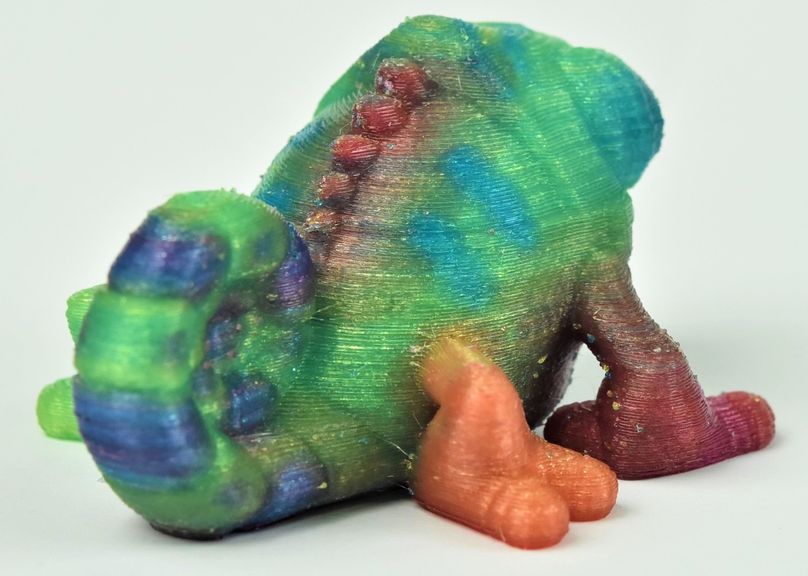}
			\end{overpic}
		\end{subfigure}
		\caption{3D printed chameleon (Thing: 2303679).
			\textbf{Top:} two views of the 3D painted model in our tool. 
			\textbf{Bottom:} photographs of the 3D printed model using our technique with matching viewpoints. The slight visual striping
			is due to varying highlights (zoom in to see the effect).} 
		\label{fig:chameleon}
	\end{center}
\end{figure}

The chameleon in Figure~\ref{fig:chameleon} has large regions of gradients and uses a variety of colors. The gradients are smooth and match the painted field closely, for example along the body around the legs. 

\begin{figure}[tbh]
	\centering
	\begin{center}
		\includegraphics[width=0.99\linewidth]{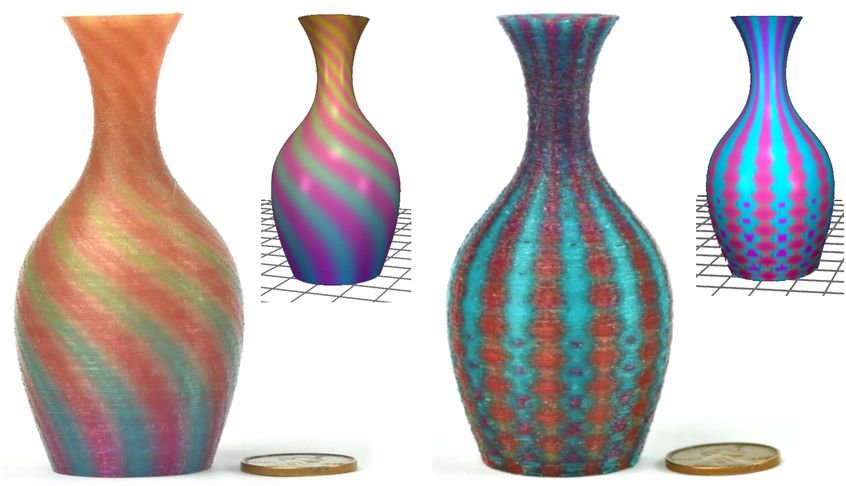}
		\caption{3D printed vases (Thing: 763726). Pictures of the print and 3D views in our paint tool in the insets, with matching viewpoints. }
		\label{fig:vases}
	\end{center}
\end{figure}

The vases in Figure~\ref{fig:vases} are examples of having both global gradients (background) and detailed patterns (stripes and dots). Both patterns and gradients are well reproduced.
In these cases the mixture fields are produced from procedural functions.


\minor{
	In Figure 18 we use an actual picture as a source for mixing ratios. Without color separation and tightly controlled filament pigmentation, we cannot currently achieve the ultimate goal of printing an accurate reproduction, or even something close to it. Instead, the user chose to create a stylized image. The resulting 3D printed lithophane exhibits little artifacts and a rich, colorfull appearance.
}
\begin{figure}[htb]
	\includegraphics[width=0.99\linewidth]{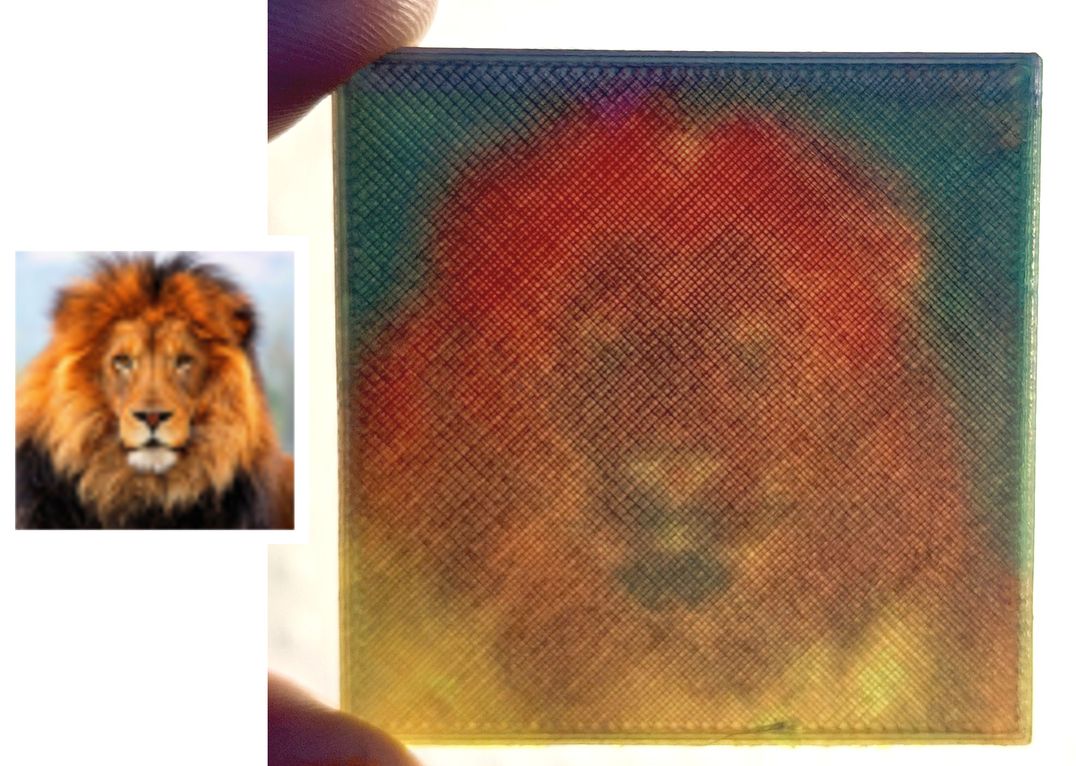}
	\caption{\minor{A low-resolution lion image is used to drive a mixing field and produce a colorfull lithophane (printed flat and backlit).}}
	\label{fig:lion}
\end{figure}

~\\ \noindent \textbf{Benefits of optimization. }
\revised{
	Table~\ref{tbl:times} summarizes timings for our main results.
	As can be seen, the optimization reduces the number of strata, and hence the print time.
	The benefits, however, depend on the complexity of the mixing field. For instance on the 
	bird model, the reduction is modest, while it is larger on the turbine model.
	This is explained by the fact that many turbine layers can be captured with only two strata.}

\begin{table}[tb]\footnotesize
	\begin{center}
		\setlength{\tabcolsep}{2pt} 
		\begin{tabular}{cccccc}
			\toprule
			Model & Mixtures & Ordering       & Est. print time & \# Layers & \# Strata      \\
			&          &                & w/ (w/o) opt.   &           & w/ (w/o) opt.  \\
			\midrule
			Dragon     & 1.9s & 6 ms & 9h20m (11h52m) & 140 & 348 (423) \\
			Chameleon  & 2.8s & 10 ms & 9h17m (10h11m) & 130 & 333 (394) \\
			Police car & 2.0s & 2 ms & 1h56m (2h19m) & 40 & 123 (164) \\
			Turbine    & 0.81s & 1 ms & 2h29m (2h32m) & 57 & 124 (174) \\
			Bird       & 3.6s & 5 ms & 6h30m (6h48m) & 239 & 651 (720) \\
			\bottomrule
		\end{tabular}
	\end{center}
	\caption{This table shows, for our main results: the time to optimize for per-strata mixtures, the time to optimize for the ordering of strata, the estimated print time for the optimized and unoptimized versions (in parenthesis), the number of layers, the total number of strata for the optimized and unoptimized versions (in parenthesis). Note that the estimated print time is underestimating the actual print time as it ignores acceleration effects.}
	\label{tbl:times}
\end{table}

\revised{
	Even when optimizing does not impact print time, it does impact quality by reducing the
	stripe defects.
	Figure~\ref{fig:cat} shows a rendering of the strata obtained with and without the optimizer.
	The number of strata goes down from 330 to 267. After optimization the mixes of the strata closely 
	match those used in the layers, whereas without the stripes of the base filaments are obvious.
}

~\\ \noindent \textbf{Towards full color printing. }
\revised{
	We tested our approach using four filaments, in particular red, green, blue and white. All the prints
	using four or five filaments use a layer thickness of $0.4$ mm.
}

\revised{
	Please note that these tests are meant to demonstrate the future potential of the technique -- accurate 
	color reproduction will require solving additional challenges, namely the color separation problem and the design
	of filaments with precisely controlled pigmentations. We are currently limited to filaments available
	from resellers, and thus cannot precisely adjust the base colors. 
}

\begin{figure}[b]
	\begin{center}
		\hspace{2mm}
		\includegraphics[height=3.2cm]{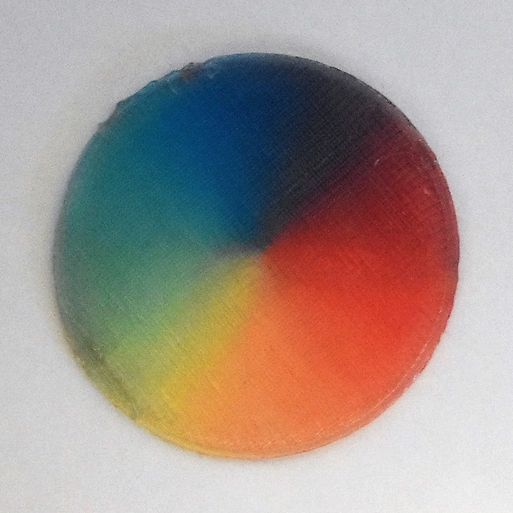}
		\hspace{3mm}
		\includegraphics[height=3.2cm]{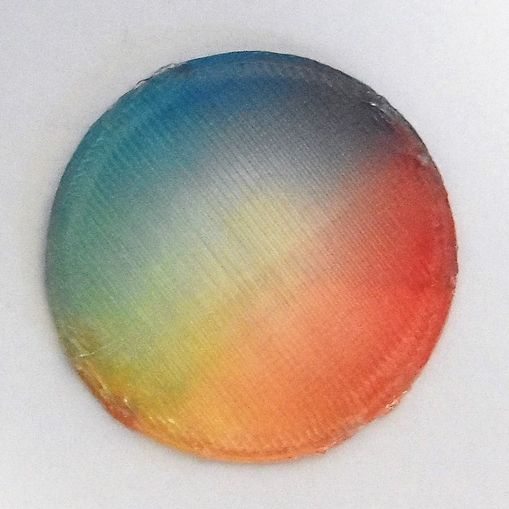}
		\caption{\revised{3D printed color discs. The left disc uses only red, green, blue mapped onto the disc radially, showing varying hues. The right disc adds white in the center, progressively changing saturation.}}
		\label{fig:colordiscs}
	\end{center}
\end{figure}

\revised{
	Despite these issues, using four filaments we can significantly increase the range of colors, as 
	demonstrated in Figure~\ref{fig:colordiscs}. 
}

\begin{figure}[t]\centering
	\includegraphics[height=3.1cm]{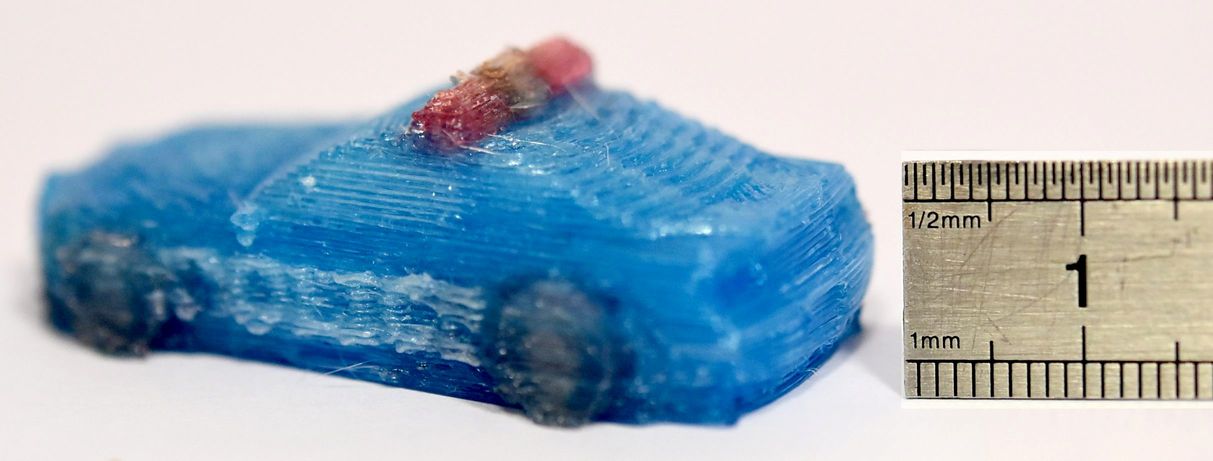}
	\caption{\revised{A tiny police car printed on a four filaments printer (Thing: 1587558 by Vorpalia). }}
	\label{fig:police}
\end{figure}

\revised{
	Figure~\ref{fig:police} shows a car printed using the same four filaments. The gray wheels are obtained
	by mixing red, green, blue, while the white is used on the sides and top.
}

\begin{figure}[tbh]\centering
	\includegraphics[height=3.1cm]{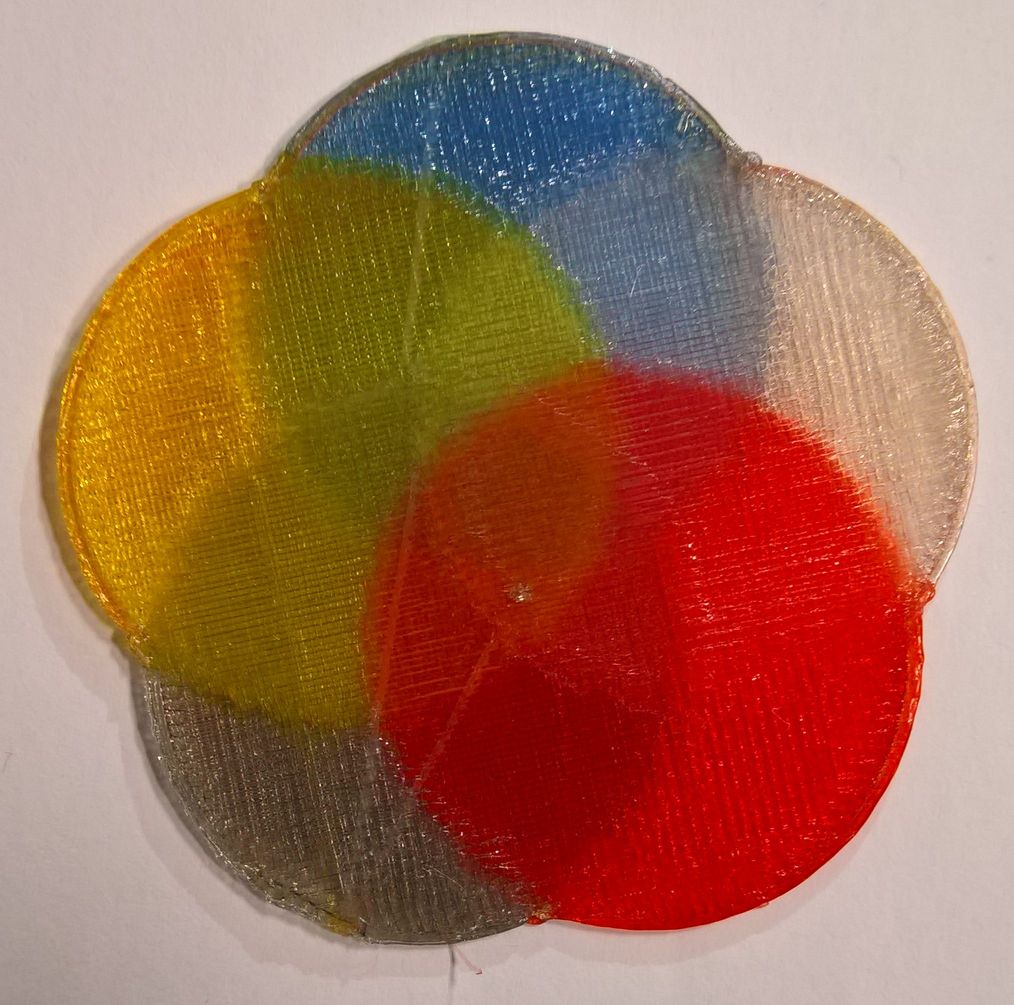}
	\includegraphics[height=3.1cm]{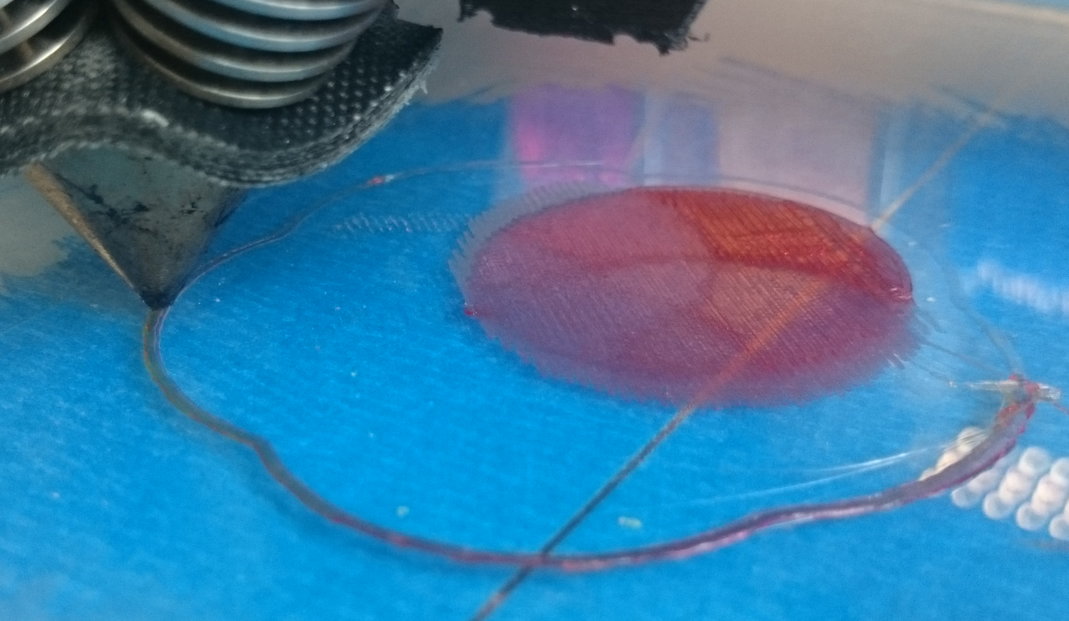}
	\caption{\revised{Five discs using each one base filament intersect in this model. This demonstrates the feasibility of using five strata with $0.4$ mm layers. The right image shows the first stratum of the first layer just after printing. The different areas corresponding to different ratios of red can be seen. }}
	\label{fig:5d}
\end{figure}

\begin{figure}[tb]
	\begin{center}
		\includegraphics[height=4.0cm]{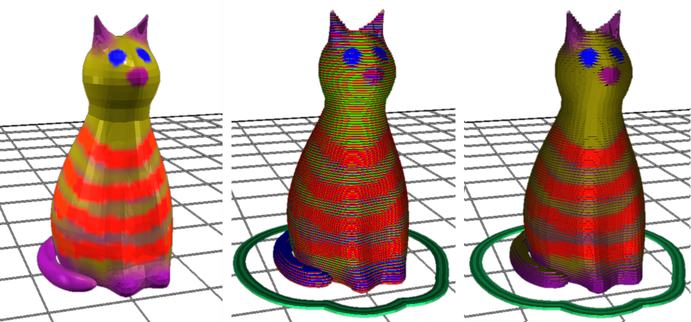}
		\caption{\revised{Painted cat using three filaments (Thing: 24255 by Mere). 
				\textbf{Left:}  3D painted model. 
				\textbf{Middle:} Strata obtained without any optimization. 
				\textbf{Right:} Strata obtained with the optimizer.}
		}
		\label{fig:cat}
	\end{center}
\end{figure}

\revised{
	Finally, we tested our technique with up to five filaments, to demonstrate the feasibility of having 
	five strata per layer, when required. This is shown in Figure~\ref{fig:5d}.
}

\section{Discussion, limitations and future work}
\label{sec:discussion}

We now discuss some specific aspects of our technique, limitations, and remaining challenges.


~\\ \noindent \textbf{Print time. }
\revised{
	One obvious drawback of our technique is that it increases print time significantly, compared to a standard print.
	At most we visit each layer once per strata, thus one can expect a proportional increase in print time.
}

\revised{
	In practice, this is mitigated by two factors. First, many strata take a zero thickness in large areas,
	allowing for fast forwarding to the next toolpaths. Second, our optimizer reduces the number of strata significantly (see Figure~\ref{fig:cat}). 
	At the extreme, if a model is painted with a constant mixture, the print time remains unchanged (a single stratum per layer suffices). 
}

~\\ \noindent \textbf{Print quality and reliability. }

\minor{
	As can be seen throughout our Figures, our prints are good but not perfect.
	Beyond the traditional defects of filament printing (z-scars, stringing), the mixture quality also suffers from oozing and smears due to the nozzle catching some material on the stratum below. These defects produce slight striations on some prints (e.g. the left vase in Figure~\ref{fig:vases}), as well as spurious material deposition (e.g. along the edges of the discs in Figure~\ref{fig:colordiscs}). 
}

\minor{The choice of filament and the printer calibration have a direct impact on these defects. The filament we used here is quite shiny and tends to emphasize the defects (see Figure~\ref{fig:chameleon}). Our printers are custom assembled from inexpensive off-the-shelves components (each printer costs around \$600), and while we carefully set them up they have limited mechanical accuracy.
}


Printing using our technique requires careful calibration and setup. The parameters have to be fine-tuned, in particular those
impacting oozing: linear advance, retraction settings, print temperature. We provide all setting and scripts for 
our slicing software on \url{https://icesl.loria.fr/pages/features/color-mixing/}.

~\\ \noindent \textbf{Mixing field complexity.}

There are limitations to the complexity of the mixture fields that can be reproduced.
Some fields may produce many isolated, tiny paths which are difficult to print reliably. 
A dedicated filtering approach would be required to remove (or enlarge) these.

\minor{
	The mixing field may also contain regions which each use a different subspace of mixtures. Our current algorithm will attempt a tradeoff across each entire layer. 
	An alternative would be to first cluster mixture ratios, and then optimize each cluster independently for base mixtures.
}

~\\ \noindent \textbf{Layer thickness. }
\revised{
	Our technique is easier to use when layers are thicker -- essentially making printing more robust to calibration errors.
	We use layer thicknesses of 0.3 and 0.4 mm. Using thicker layers would be possible, but we are limited by the nozzle 
	output diameter (0.4 mm ; it is generally advised to use layer thicknesses below the nozzle diameter).
	In principle, our technique works with adaptive layer thicknesses, but we have yet to experiment with this. Of course,
	the range of adaptivity is limited by the thinnest achievable layers.
}


~\\ \noindent \textbf{Nozzle shape and up/down z motions. }
Abrupt changes in Z (height) coordinates while printing paths can produce defects: the flat area at the nozzle tip interferes with already printed paths around. Song et al. \shortcite{Song:2017} address this issue for printing top surfaces with better accuracy, splitting and re-ordering paths to minimize interference. However, our case is less sensitive to this issue. While the first strata are indeed curved, the last stratum always exactly aligns with the (flat) layer top boundary. As a consequence, the last stratum does not suffer from nozzle interferences (see Figure~\ref{fig:teaser}, rightmost). While there may be small geometric defects in between strata, we found that in practice they have no impact on the final perceived quality, especially as the correct amount of (translucent) colored filament was deposited.

~\\ \noindent \textbf{Filament availability. }

The results of our technique depend on the source filament quality. In particular, it is important for the filament pigmentation to be well balanced. Otherwise, even tiny amounts of a filament with high pigment concentration will immediately overrule the other ones.

\minor{
	We found it very difficult to source well suited filament from existing vendors. We now produce our own filament using an off-the-shelf extruding machine designed for producing 3D printer filaments. We mix pigments with PLA pellets prior to feeding them to the machine. We use cyan, magenta and yellow pigments. We provide more details on \url{https://icesl.loria.fr/pages/features/color-mixing/}.
}




\section{Conclusion}

We introduced a technique for colored printing with controlled gradients using
readily available low-cost printer components and filaments. 
Beyond the initial idea of mixing filaments through the use of per-layer strata, 
we propose an optimization of the mixture of each stratum to improve the quality and
reduce print time, exploiting a mixing nozzle.
One important advantage of our technique is that it keeps using standard materials
(PVA, ABS, PET, etc.) and does not impose one particular type of filament.

As the technologies evolve we believe our approach will remain useful: by using 
constant color strata we improve deposition continuity. With more precise
hardware and controllers, oozing could be reduced and higher quality deposition 
would be achieved.
%
An interesting question is whether our approach could be used to grade properties
beyond colors, such as elasticity using flexible filaments.


\begin{acks}
This work was funded by ERC ShapeForge (StG-2012-307877) and received support
from R\'{e}gion Lorraine and FEDER. We thank C\'{e}dric Zanni for his help with
preparing colored prints, as well as Sandrine Hoppe (LRGP) for 
discussions regarding filaments and materials.
\end{acks}


\citestyle{acmauthoryear}
\setcitestyle{square}
\bibliographystyle{ACM-Reference-Format}
\bibliography{paper}

\end{document}